# On Locality, Growth and Transport of Entanglement

## Roland Omnès

*Abstract*

        Entanglement of a macroscopic system with a microscopic one is shown to begin by a topological property of histories in the Feynman formulation of quantum mechanics. This property can also be expressed algebraically on the Schrödinger equation through a convenient extension of the Hilbert space formalism. Entanglement shows then properties of growth and transport, the corresponding local and temporary character of entanglement being called here "intricacy" when it occurs. When applied to the continuous interaction of a macroscopic system with a random environment, intricacy implies a "predecoherence" effect, which can generate and transport permanently incoherence within the system. The possible relevance of these results for a theory of wave function collapse is also indicated.

R. Omnès

Laboratoire de Physique Théorique (Unité Mixte de Recherche (CNRS) UMR 8627),

Université de Paris XI, Bâtiment 210, 91405 Orsay Cedex, Francee

e-mail: Roland.Omnes@th.u-psu

————————

## 1. Introduction

        When Schrödinger introduced the notion of entanglement [1] (see also [2]), he took as a paradigm the case of two quantum systems $S$ and $S'$, which are both initially in a pure state, then come to interact and finally separate to become again isolated. Although the compound system $S + S'$ is still again in a final pure state, this is no more true separately of $S$ or $S'$ and each one of them appears in a mixed state when considered by itself. Schrödinger, who regarded entanglement as the main character of quantum mechanics, showed also in a series of famous papers [3] how it makes a quantum description of a measurement incompatible with the uniqueness of macroscopic data: When $S$ is a measuring apparatus and $S'$ a measured system, the Schrödinger equation is deterministic and different values of a measured observable $Z$ determine differently the dynamics of interaction. When the initial state of $S'$ is a superposition of different eigenvectors of $Z$, entanglement separates forever as many different states of $S$ with no communication [4], bringing out an essential difficulty for understanding how an actual unique datum can arise from an actual individual measurement.

        In the present work, one studies entanglement as a local and evolving property of a macroscopic system $S$, particularly when it interacts with a microscopic system $S'$ (for instance a particle). This local aspect of entanglement was not ignored previously and it was studied particularly from the standpoint of relative entropy, in search for a measure of entanglement [5], but the present approach is different and leads more easily to wider results. Some refinement in vocabulary is however convenient to distinguish the corresponding local and transitory effects from a final full and definitive entanglement and one will therefore call

them "intricacy" for convenience: In that sense, intricacy will appear as a special set of properties, which characterize entanglement locally in the course of time.

In Section 2, intricacy is envisioned in the framework of Feynman histories and its topological character is pointed out. An algebraic formulation is then constructed in Section 3, showing that intricacy is not a standard physical property as defined by von Neumann (it cannot be associated with a projection operator in Hilbert space [4]). Nevertheless, a measure for intricacy of an individual atom in the macroscopic system $S$ is obtained. This is extended in Section 4 to the expression of an average amount of intricacy in a finite region of space at a given time, by means of a more powerful approach using quantum field theory and showing a rather wide generality of intricacy.

Section 5 deals with the transport properties of intricacy, whose measure evolves as a nonlinear wave with a finite velocity, which can be the velocity of sound, the Fermi velocity of electrons in a conductor or the velocity of light, according to the agents carrying intricacy in various conditions.

Section 6 is devoted to an examination of various cases of entanglement and intricacy, especially in quantum measurements, with a possible application to the problem of wave function collapse.

## 2. Intricacy of an atom in Feynman histories

One will use mainly an example where the macroscopic system $S$ consists of a gas of $N$ neutral atoms at standard temperature $T$, enclosed in a box. The system $S'$, which is microscopic, is an energetic particle $M$ going along a straight-line trajectory and interacting with nearby atoms when it crosses the box. After separation, the two systems $S$ and $M$ are entangled.

The evolution of $S$ after separation can be described in terms of Feynman histories. A Feynman history consists then of a path for $M$ and individual paths for the $N$ atoms, or equivalently a path of $N$ atoms in the configuration space of the gas. A specific history starts from a definite set of positions $\{x_0\}$ for the atoms and a position $y_0$ for $M$, at a time $0$ prior to any interaction between $M$ and $S$. Later on, the positions of the atoms and of $M$ vary arbitrarily with time as long as $M$ crosses the box, and one considers a still later time $t$ after the $M$-$S$ separation. The amplitude of a history $h$, ending with positions $\{x\}$ for the atoms and $y$ for particle $M$ at time $t$, is given then by

$$A_h(\{x_0\},y_0;\{x\},y;t) = K \exp[i\,\Sigma(\{x_0\},y_0,0;\{x\},y,t)/\hbar\,],  \qquad (2.1)$$

where $K$ is a normalization factor and $\Sigma$ is the classical actions along history $h$.

One will assume that the interaction potentials between two atoms have a short finite range $b$, so that a contribution of the potentials to the action $\Sigma$ occurs only when two individual atom paths come at a distance shorter than $b$ at some intermediate time. To make the discussion simpler, one will also take a model where the interaction of the particle $M$ with an atom is due to a potential with a finite range $b'$. One will then say that a specific atom $a$ is intricate or 'connected' with $M$ at time $t$ in this history when there is at least one chain of interactions linking $M$ with $a$, either from a direct $a$-$M$ interaction or from a 'contagion' of intricacy, which occurred when there was an interaction of $a$ with a previously intricate atom *a'*. Accordingly, the intricacy of *a* with *M* can be either due to a direct interaction with *M* or caught from contagion.

This is clearly a topological property of the Feynman graph representing the history and one can split the set of histories into two subsets: Some histories, denoted by an index (*a*1), involve a connection of *M* with *a* whereas other histories, denoted by an index (*a*0), show no such connection. This splitting can be extended to an *M-S* wave function $\Psi$ through the relation

$$\Psi_{a\mu}(\{x\},y;t) = \int \{dx_0\} dy_0 \Psi_0(\{x_0\},y_0) \sum_{h_\mu} A_{h_\mu}(\{x_0\},y_0;\{x\},y;t) \qquad (2.2)$$

where $\mu$ is an intricacy index for atom *a* with the value 0 or 1 and $\Psi_0(\{x_0\},y_0)$ is the wave function at time 0. The summation is performed over histories in which the atom *a* became intricate with *M* before time *t*, in which case the intricacy index is $\mu = 1$, or *a* is still not intricate, in which case $\mu = 0$. The wave function $\Psi$ is thus a sum

$$\Psi(\{x\},y;t) = \Psi_{a0}(\{x\},y;t) + \Psi_{a1}(\{x\},y;t). \qquad (2.3)$$

This intricacy of an individual atom referring to Feynman histories is accordingly a topological property of connection arising from a contagion, either due to a direct interaction with *M* or caught from previously intricate atoms and eventually transmitted to other atoms. This notion of intricacy for an atom can be also extended to a subset of the set of atoms in *S* and one will show later on how it can be defined also for the atoms in some space region within the box enclosing *S* at time *t*.

A difference appears thus between entanglement and intricacy: As introduced by Schrödinger, entanglement refers to the whole system *S* and includes all the atoms in it. It is established as soon as the two systems *M* and *S* separate. Intricacy is more detailed and grows through contagion after separation, until all atomic states in *S* have become intricate. Only then does intricacy coincide with entanglement.

## 3. An algebraic expression of intricacy

One considers now the contagion of intricacy in relation with Schrödinger's equation. One assumes that the state of particle *M* is initially given by a wave function $\chi(y)$. The state of the macroscopic system *S* is unavoidably mixed and represented by a density matrix $\rho_S$ with eigenfunctions $\psi(\{x\})$. At time 0, before interaction, a *M-S* wave function is therefore a product

$$\Psi(\{x\},y;0) = \psi(\{x\};0)\chi(y,0). \qquad (3.1)$$

Its evolution is given by the Schrödinger equation

$$i\hbar d\Psi/dt = (K_M + K_S + U + V)\Psi, \qquad (3.2)$$

where $K_M$ and $K_S$ are the kinetic energies of the particle *M* and of the atoms in the gas. The potential *U* is a sum of interactions of *M* with the various atoms, and *V* is the sum of potential interactions between atoms. Denoting these atoms by an index *n*, one has then:

$$U = \sum_n U(y,x_n), \qquad V = \sum_{(n,n')} V_{nn'}(x_n,x_{n'}), \qquad (3.3)$$

where the summation in the last expression is performed over pairs of atoms.

There is in principle no difficulty in following the development of interactions and the corresponding growth in intricacy during the period when particle $M$ is crossing the gas but, for brevity, one will concentrate on the later period when $M$ and $S$ have become again separated and the growth of intricacy is pure contagion. The density matrix $\rho_S$ evolves then unitarily under the Hamiltonian $K_s + V$ and one still denotes by $\psi(\{x\})$ one of its eigenfunctions.

One introduces again, for every atom $n$, a pair of indices $\mu = 0$ or 1 according to its intricacy or lack of intricacy with $M$ at some time $t$. A detailed expression of intricacy for all the atomic states is therefore specified by a string $q$ of $N$ indices 0 or 1. The previous use of Feynman histories allow one to write down formally a wave function $\psi(\{x\}, t)$ as a sum of functions $\psi_q(\{x\}, t)$, corresponding to different connections in underlying Feynman histories:

$$\psi(\{x\}, t) = \sum_q \psi_q(\{x\}, t). \qquad (3.4)$$

For each atom with index $n$, one introduces then several operations regarding its two types of intricacy. They are written as $2 \times 2$ matrices acting on a two-dimensional space in which the two basis vectors are characterized by the intricacy indices 0 and 1 for atom $n$. The corresponding algebra is

$$P_{\mu n}{}^2 = P_{\mu n}, \qquad P_{1n} + P_{0n} = I_n \; ; \qquad P_{1n}P_{0n} = P_{0n}P_{1n} = 0,$$
$$S_n{}^2 = 0, \qquad S_n P_{0n} = S_n, \quad P_{0n}S_n = 0, \quad S_n P_{1n} = 0, \qquad P_{1n}S_n = S_n \qquad (3.5)$$

$P_{0n}$, for instance, is an operation leaving unchanged a function $\psi_q$ in which the index $n$ is 0, and annihilating $\psi_q$ if this index is 1. It is therefore a projection operator for intricacy 0 or 1. $I_n$ is the identity matrix, which does not affect intricacy. $S_n$ is an operation acting on the intricacy index 0 and replacing it by index 1. No inverse operation going from intricacy 1 to no intricacy 0 is introduced, because a topological connection is established once for all in a Feynman graph. Leaving aside the index $n$, these operations for atom $n$ can be expressed explicitly in terms of Pauli matrices by

$$P_0 = (I - \sigma_z)/2, \; P_1 = (I + \sigma_z)/2, \; S = \sigma_+ = (\sigma_x + i\sigma_y)/2.) \qquad (3.6)$$

One can then obtain simply a transition from the topological properties of Feynman graphs to algebraic properties of indexed wave functions, as follows: One introduces a large linear space $E'$, which consists of $2^N$ copies of the space $E$ of wave functions $\psi$. Each copy corresponds to a definite string $q$ and this splitting of $E$ into copies corresponds then to an expansion (3.4) showing intricacy for some Schrödinger wave function $\psi$. The different copies of $E$ are not mutually orthogonal, since two functions $\psi_q$ in (3.4) need not be orthogonal, so that $E'$ is a linear space but not a Hilbert space. The standard Hilbert space $E$ is on the other hand a projection of $E'$, as shown by (3.4). Such a correspondence is typical of topological properties and is encountered for instance in Riemann surfaces for multi-valued analytic functions over the complex plane. It is usually expressed in the framework of fiber bundles [6], but one will not enter into these refinements.

One also introduces an extended Hamiltonian operator $H'$, which acts in the space $E'$; according to

$$H' = \sum_n -I_n(\hbar^2/2m)\nabla_n^2 + \sum_{(n,n')} V(x_n, x_{n'}) O_{nn'} \tag{3.7}$$

where. $O_{nn'} = P_{0n} \otimes P_{0n'} + P_{1n} \otimes P_{1n'} + S_n P_{0n} \otimes P_{1n'} + P_{1n} \otimes S_{n'} P_{0n'}.$ (3.8)

The operator $O_{nn'}$ retains the essential character of intricacy, namely: When two atoms $n$ and $n'$ are both intricate or are both non-intricate, there is no change in their common intricacy when they interact. If on the other hand, one of them is initially intricate and the other one is not, both of them become intricate after interaction. In both cases anyway, the interaction is still due to the potential $V$ and no new dynamical effect is introduced. There is only a contagion of intricacy in which one atom can catch intricacy from a collision with a previously intricate atom and can then transmit it to other atoms, as the connections in Feynman histories.

One can take as an example the case of a simple system $S$ consisting of three atoms $a$, $b$, $c$: There are eight functions $\psi_q(x_a, x_b, x_c)$ with eight evolution equations. One of these equations, in the case of $q = (011)$, is for instance

$$i\hbar(\partial/\partial t)\psi_{011} = -(\hbar^2/2m)(\nabla_a^2 + \nabla_b^2 + \nabla_c^2)\psi_{011}$$
$$V(x_a, x_b)\psi_{001} + V(x_a, x_c)\psi_{010} + V(x_b, x_c)\psi_{011}. \tag{3.9}$$

One notices that the Hamiltonian $H'$ is not a self-adjoint operator in $E'$, as a reflection of the fact that $E'$ is not a Hilbert space, so that scalar products or properties of adjointness are valid only in the Hilbert space $E$ of standard wave functions $\psi(x_a, x_b, x_c)$.

This remark is significant, because it points out why intricacy is not a standard quantum property: According to Von Neumann [4], a quantum property should always be associated with a projection operator in the Hilbert space $E$ and stands itself as an observable. Although every string of intricacy could be associated formally with a projection operator acting on the extended space $E'$, this operator does not act on $E$ and is not an observable.

Another difference between intricacy and ordinary physical properties can be seen also in the case of a gas: Two-particle scattering is a physical effect and the collision matrix can be described in perturbation theory by a sum over all the terms of a perturbation expansion [7]. But the Hamiltonian $H'$ in (3.7) implies that intricacy for atom $n$ is completely established by the first factor $V$ occurring in any term of a perturbation expansion. This non-gradual behavior is therefore again an expression of the non-gradual character of a topological property.

The consistency of this approach with basic quantum mechanics results from a simple algebraic property of the $2^N$ evolution equations generalizing (3.9): If one applies the operator

$$A = \prod_n (P_{1n} + S_n P_{0n}) \tag{3.10}$$

to a function $\psi_q$, one obtains the same function of the atomic positions $\{x\}$, but now associated with the string of complete intricacy $Q = (11111...1)$. When multiplying an operation $O_{nn'}$, this operation $A$ has also the property of bringing the intricacy indices of the two atoms $(n, n')$ on the string (11). As a result, when one applies the operation $A$ on the two sides of the extended Schrödinger equation

$$i\hbar(\partial/\partial t)\sum_q \psi_q(\{x\}, t) = \sum_q H' \psi_q(\{x\}, t), \tag{3.11}$$

one obtains the standard Schrödinger equation for the sum (3.4), namely

$$i\hbar \partial \psi / \partial t = (K_S + V)\psi. \tag{3.12}$$

One may notice that this agreement with the standard Schrödinger equation is obtained by means of a method in which the wave function $\psi(\{x\}, t)$ is forced to become associated with the string (1111...) of complete intricacy. This is in close agreement with the fact that Hilbert space methods can only account for entanglement and not for the topological details of intricacy.

As a last remark about this construction, one may look at the question of giving a measure to intricacy [5]: It would be tempting to write down explicit probabilities $p_{a1}$ and $p_{a0}$ for some atom $a$ (or some subset of atoms) to be intricate or not with $M$ at some time $t$. One can use for this purpose an ordering in the strings of intricacy indices so that the intricacy index of an atom $a$ is written as the first bit, the strings in which $a$ is intricate with $M$ having then a form $(1q')$, where $q'$ is an arbitrary string of $N-1$ atoms. One can then define two quantities

$$p_{a1} = \left\| \sum_{q'} \psi_{1q'} \right\|^2 \text{ and } p_{a0} = \left\| \sum_{q'} \psi_{0q'} \right\|^2. \tag{3.13}$$

as candidates for atomic measures of intricacy. The sum of these expressions differs however from 1 by the quantity

$$2\text{Re}\left( \left\langle \sum_{q'} \psi_{0q'} \middle| \sum_{q''} \psi_{1q''} \right\rangle \right). \tag{3.14}$$

This difference is certainly negligible when summed over all the eigenfunctions of $\rho_S$ on which one must sum up ultimately, because of the complexity of these wave functions for a macroscopic system and of their huge number. One can then conclude by saying that $p_{a1}$ is a reasonable measure of intricacy for the atom $a$, but the limited scope of this result indicates that much more is needed and the theory must be still widened, as one will do now.

## 4. Local intricacy

Two points in the previous discussion need obviously improvement. The first one laid in the consideration of individual atoms, although the atoms in a gas are generally undistinguishable. The second point, closely linked with the first one, was that knowing the probability of intricacy for a specific atom is neither of much practical nor of conceptual interest. One would be much more interested in local properties of intricacy and, for instance, one would like to know a measure of intricacy for all the atoms in some region of space within a detector. One expects that, when a charged particle $M$ has just crossed rapidly the gas, there is much more intricacy near the track of $M$ than farther away. On the other hand, the growth of connection with time in Feynman histories would imply that all the gas becomes gradually more intricate with $M$ as time goes on.

All these questions are directly concerned with locality and they become much clearer when the framework of quantum field theory is used. To begin with, one will still work with the previous non-relativistic example, in which case field theory amounts to second quantization [8]. The atoms are described by a field $\varphi(x)$ where the notation $x$ involves not only the position of an atom, but also spin indices. The field satisfies either commutation or anti-commutation properties according to the spin value, but the two cases are very similar and one will retain only for illustration the case of Bose-Einstein statistics, with commutation relations

$$[\varphi(x),\varphi(x')] = 0, \qquad [\varphi(x),\varphi^\dagger(x')] = \delta(x - x').\tag{4.1}$$

When considering the gas, one must introduce a vacuum state $|0\rangle$. A state of the gas with $N$ atoms and a wave function $\psi(\{x\})$ is then given by

$$|\psi\rangle = \int\{dx\}\psi(\{x\})\prod_{j=1}^{N}\varphi^\dagger(x_j)|0\rangle.\tag{4.2}$$

The Hamiltonian is

$$H = \int dx\varphi^\dagger(x)(-\nabla^2/2m)\varphi(x) + (1/2)\int dxdx'\varphi^\dagger(x)\varphi^\dagger(x')\varphi^\dagger(x)V(x,x')\varphi(x)\varphi(x').\tag{4.3}$$

(The factor 1/2 in the last term is due to the fact that a pair of atoms with positions $x$ and $x'$ occurs twice in this expression, in the orderings $(x, x')$ and $(x', x)$). The local density of atoms is given by $\rho(x) = \varphi^\dagger(x)\varphi(x)$

The evolution of intricacy under connection, as discussed in Section 3, can also be expressed in field formalism if one introduces an intricate field $\varphi_1(x)$, associated with intricate states of atoms, and a non-intricate field $\varphi_0(x)$. The commutators $[\varphi_0(x),\varphi_0^\dagger(x')]$ and $[\varphi_1(x),\varphi_1^\dagger(x')]$ are still given by (4.1) whereas $\varphi_0(x)$ and $\varphi_0^\dagger(x)$ commute with $\varphi_1(x')$ and $\varphi_1^\dagger(x')$. A linear space $E'$ describing intricacy over the Hilbert space $E$ of physical states can again be introduced as in (3.4). In the case of three atoms and of the string $q= 011$ for instance, one would write a state with intricacies $(0, 1, 1)$ as

$$|\psi_{011}\rangle = \int dx_1 dx_2 dx_3\psi_{011}(x_1,x_2,x_3)\varphi_0^\dagger(x_1)\varphi_1^\dagger(x_2)\varphi_1^\dagger(x_3)|0\rangle.\tag{4.4}$$

Because of the commutative properties of fields, it is clear that this state is symmetric (or anti-symmetric) under a permutation of identically intricate atoms.

The Hamiltonian corresponding to (3.7) is then

$$\begin{aligned}
H' = &\int dx\{\varphi_0^\dagger(x)(-\nabla^2/2m)\varphi_0(x) + \int dx\{\varphi_1^\dagger(x)(-\nabla^2/2m)\varphi_1(x)\\
&+ (1/2)\int dxdx'\{\varphi_0^\dagger(x)\varphi_0^\dagger(x')V(x,x')\varphi_0(x)\varphi_0(x') + \varphi_1^\dagger(x)\varphi_1^\dagger(x')V(x,x')\varphi_1(x)\varphi_1(x')\}\\
&+ (1/2)\int dxdx'\varphi_1^\dagger(x)\varphi_1^\dagger(x')V(x,x')\{\varphi_1(x)\varphi_0(x') + \varphi_0(x)\varphi_1(x')\}
\end{aligned}\tag{4.5}$$

The field equations $i\partial\varphi/\partial t = [H',\varphi]$ become

$$i\partial\varphi_0(x,t)/\partial t = -\nabla^2\varphi_0(x)/2m + \left\{\int dx'(\varphi_0^\dagger(x')V(x',x)\varphi_0(x'))\right\}\varphi_0(x),$$

$$i\partial\varphi_1(x,t)/\partial t = -\nabla^2\varphi_1(x)/2m + \left\{\int dx'(\varphi_1^\dagger(x')V(x',x)\varphi_1(x'))\right\}(\varphi_0(x) + \varphi_1(x))$$

$$\tag{4.6}$$

When used to follow intricacy as in Schrödinger's consideration of three successive steps (separation/interaction/separation) in entanglement [1], this formalism would start during the first step of separation with a unique field $\varphi_0(x)$. If the second period involves only interactions with a particle $M$, associated with a field $\chi(y)$ and interacting with the atoms in the gas through a potential $U$, the generation of intricacy would be due to a corresponding term in the $M$-$S$ Hamiltonian:

$$H_{MS} = \int dx dy \varphi_1^\dagger(x) \chi^\dagger(y) U(x,y) \varphi_0(x) \chi(y). \tag{4.7}$$

The Hamiltonian $H'$ of $S$ in (4.5) would already contribute to the growth of intricacy during this second period when the systems $M$ and $S$ interact and it would completely control the contagion during the last period when $M$ and $S$ are again separated. It should be clear also that the simple expression (4.7) for the generation of intricacy can be extended properly to some more elaborate cases where the second system $S'$ does not consist of a unique particle $M$, but one will not try to develop this aspect here.

This approach is closer to the reality of physics than the previous one in Section 3. It yields for instance directly a simple expression for a local measure –or probability- of intricacy: If the gas is homogeneous and one denotes by $n$ the average number of atoms per unit volume, one can define average probabilities for intricacy (or no-intricacy) in a macroscopic space region $V$ by

$$p_\mu(V) = (1/nV) Tr \left\{ \int_V \rho_\mu(x) dx \right\}, \tag{4.8}$$

with $\rho_\mu = \varphi_\mu^\dagger(x) \varphi_\mu(x)$. The sum of these two quantities for $\mu = 0$ and 1 is certainly very close to 1, for the same reasons as above.

A great power of the field approach is its flexibility. It applies to all kinds of particles, either fermions or bosons, to systems containing a unique kind of atom or different kinds, to relativistic or non-relativistic dynamics. The method works when all the atoms are in their ground state, but also when there are excited states: The various states of excitation can then be associated with different fields. When ions and electrons are produced through interactions with incoming charged particles, one may use an explicit or implicit framework involving the quantum fields of nuclei, electrons and photons. In a solid system or subsystem, one can use other fields to describe phonons or conducting electrons. In that sense, one may presume that this approach, in which intricacy is considered as a topological connection of Feynman histories for quantum fields, has a wide domain of validity and a significant meaning.

## 5. Growth and transport of intricacy

One thus arrives at a representation of intricacy as a topological property of the history of atomic states, carried by the motion of atoms and transmitted by them through interactions, particularly under collisions. This conception is strongly reminiscent of a transport process (like heat conduction or conductivity for instance) and one will therefore look now at intricacy from this standpoint.

A significant difference between intricacy and other transport processes is however its contagious character, which is in opposition with the conservation of energy or or of electric charge, for instance. There is an accumulation in the amount of topological connection rather than a share of a conserved physical quantity. The image of a contagion suggests nevertheless a local growth of intricacy at the place where it has been created (near the track of a charged particle for instance) together with a transport from a rich region where a large fraction of states are intricate towards poor regions where this fraction is lower. Ultimately, after enough time, intricacy will be total everywhere and valid for the whole state of an entangled system.

One needs some measure of intricacy to give a substance to this representation, and this measure must be local. To define it, one will use Equation (4.8) and denote by $f_1(x, t)$ the average probability for intricacy of the atomic states in a small macroscopic region around a space point $x$ at time $t$. Similarly, $f_0(x, t)$ will denote the corresponding measure for no-intricacy and one will assume again that the sum $f_0(x, t) + f_1(x, t)$ is very close to 1.

The transition from quantum elementary effects to the kinetic behavior at a larger scale is however among the trickiest points in theoretical physics. One must always rely ultimately on a model of classical atoms (or other carriers) undergoing a random motion [9], whereas average exchanges in the transported quantity are derived from quantum mechanics (through scattering theory for instance [7]). Here, although intricacy is an extreme paradigm of quantum properties (since it is not even associated with an observable), its exchange properties are extremely simple: Both atoms in the final state of a collision have intricacy 1 if one of them has initially this intricacy. One will therefore use a standard kinetic method in the present case.

One will take the case when the gas in the previous model is at thermal equilibrium with temperature $T$. The average velocity of an atom is then $v = (3k_B T/2m)^{1/2}$ and one also introduces its mean free time $\tau$ and its mean free path $\lambda = v\tau$. One considers then that the motion of atoms is a random walk. If the intricacy of an atom were conserved in a collision, intricacy would be a conserved quantity and the random walk of an atom with intricacy 1 would arise from collisions with other atoms, with no difference when these atoms have intricacy 0 or 1. The evolution of the quantity $f_1(x, t)$ would then be simply a diffusion effect, governed by the equation

$$\left(\partial f_1/\partial t\right)_{diffusion} = (1/6)\nabla^2 f_1, \tag{5.1}$$

where the units of length, time (and velocity) are taken as $\lambda$, $\tau$ (and $v$). The diffusion coefficient 1/6 results then from the random walk of atoms in three-dimensional space.

In the present case, there is in addition a local growth in intricacy. Its probability of occurrence during a time interval $\tau$ is the product of the probability $f_1(x, t)$ for one of the atoms to be intricate by the probability $f_0(x, t)$ for the other atom to be not intricate. The corresponding increase in local intricacy is then given by

$$\left(\partial f_1/\partial t\right)_{contagion} = f_1(1 - f_1), \tag{5.2}$$

where one took account of $f_0 = 1 - f_1$..

The two variations (5.1) and (5.2) add up and one has then for the total variation:

$$\partial f_1/\partial t = f_1(1 - f_1) + (1/6)\nabla^2 f_1. \tag{5.3}$$

In principle, initial conditions for this equation are straightforward: If one wished to describe the initial generation of intricacy (for instance under the action of the previously interacting $M$ particle), one would introduce a source term in the right-hand side of (5.3) and $f_1$ would be equal to zero before interaction of $M$ with $S$. If one were however interested only in the growth and transport of intricacy after separation of $M$ and $S$, one would give some initial condition $f_1(x,0)$ at a time 0 when $S$ and $M$ separate.

*Boundary conditions*

The boundary conditions for the nonlinear partial differential equation (5.3) are not obvious, on the contrary. The domain of definition, which can depend on time, is essentially controlled by the process of intricacy transport itself, which defines the region where $f_1$ is a non-negative quantity. The boundary condition is therefore $f_1 = 0$, but on a boundary depending itself on the equation that must be solved. This is typical of some nonlinear equations showing a moving front [10], and there is such a front in the present case, as one will discuss now.

An essential condition on this equation is that it makes sense only when and where $f_1$ $(x, t)$ is non-negative. Although no rigorous analysis of (5.3) could be made, rough investigations suggested that this positivity condition would never be satisfied in infinite space, whereas it is valid for the linear diffusion equation (5.1). Rather than attempting a rigorous mathematical analysis, which seems difficult, one will therefore try another approach relying on physical considerations: The basic problem is to find the domain in which (5.3) holds together with the positivity condition. To get a hint, one thinks of a model where intricacy is generated at time 0 on a plane $z = z_0$, in equal amount at all the points in this plane. The corresponding equation would be

$$\partial f_1 / \partial t = f_1(1 - f_1) + (1/6)\partial^2 f_1 / \partial z^2 + \delta(t)\delta(z - z_0). \qquad (5.4)$$

One does not aim however at solving this equation, but to get an understanding of the phenomena it describes, with the help of their one-dimensional behavior.

There is a source of simplicity in the problem, which is not apparent in the transport equations, because of its direct origin in the quantum principles. The atoms, which catch and transmit the contagion of intricacy, are identical. When an intricate atom collides with a non-intricate one, both of them come out intricate and cannot be distinguished. In the reference system of their center of mass, one of them goes however towards the positive direction of the $z$-axis and the other one towards the negative direction and something of that trend must remain on average in the laboratory frame. Intricacy must grow and move from its plane of origin at some velocity. The velocity distribution of atoms along the $z$-direction is the one-dimensional quantity $v' = (k_B T/2)^{1/2} = 3^{-1/2}v$, and the transport of intricacy together with its continuous generation must therefore imply, at a macroscopic scale, its propagation away from the starting plane behind a plane wave front (or rather two opposite fronts), which moves at a velocity of order $v'$.

This means that the boundary condition for Equation (5.3) should be that $f_1$ vanishes on some moving front (or fronts), whose position is determined by extraneous considerations involving the distribution of atoms and originating in their indistinguishability. As a consequence, the geometry of boundaries must be found in the distribution of intricate atoms in phase space, whereas $f_1$ and $f_0$ describe only their distribution in space.

Although the problem is much involved, one can easily get for it a manageable formulation, which requires however knowledge of the wave front velocity. One knows that it is of order $v'$, but that does not yield its exact value. Different models have been tried and yield different answers, ranging from $v'$ to $(2/\pi)^{1/2}v'$ (which is the average one-dimensional velocity of atoms going in the positive $z$-direction). Rather than attempting a difficult rigorous analysis, which will be left for later work, one will use here the value $v'$, which is the velocity of sound. This is because of a strong analogy between the contagion of intricacy with a chain reaction, which can be considered also as a detonation with vanishing heat reaction and should therefore have a wave front moving at the velocity of sound (*i.e.*, a shock wave with vanishing discontinuities [11].

*The front shape*

The problem of growth and transport of intricacy becomes then remarkably simple. Intricacy begins in some initial region (the track of particle $M$ for instance). It becomes soon complete or nearly complete inside a growing region whose boundary at time $t$ is at a distance $v't$ from the track. This boundary (or wave front) is locally plane and the problem of evolution of intricacy amounts at finding the behavior of $f_1$ near this front. Letting $z$ denote a local coordinate in the normal direction to the front, one can write $f_1 = g(z - v't)$ and (5.3) becomes a simple nonlinear differential equation for $g$: as a function of $x = z - v't$:

$$-3^{-1/2}g'(x) = g(x)(1 - g(x)) + g''(x)/6 .\qquad(5.5)$$

with boundary conditions $g(0) = 0$ and $g(-\infty) = 1$, the wave front being located at $x = 0$.

One can write $g(x) = 1 - u(x)$ for $x$ large and negative, and (5.5) shows then that $u(x)$ has the asymptotic behavior $C\exp(qx)$ for large negative values of $x$, with $q = 3 - 3^{1/2}$. The choice of the constant $C$ depends on the exact distance of the front to the original track, which is unessential when this distance is large. One can then choose a point with abscissa $-x_0$ where the asymptotic exponential expression of $u$ is valid with $C = \exp(-q\,x_0)$ and Figure 1 shows $g(x)$ as obtained from a numerical computation with $C = 0.05$. (This figure could give the impression that $g'(x)$ vanishes on the front, but this is impossible. The actual result of the computation is a small number $g'(0) \approx$ -0.06.

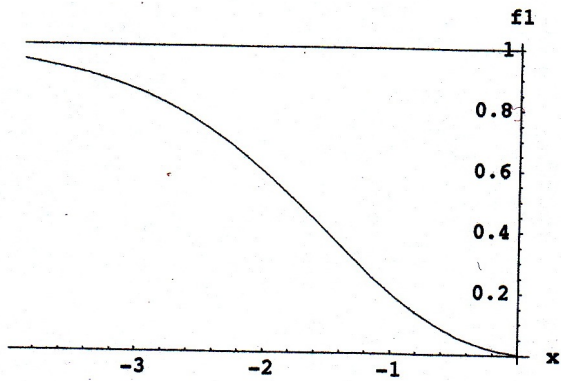

Figure 1: An intricacy wave, The local probability of intricacy behind a wave front at x = 0, with a unit of distance taken as the atomic mean free path.

*More general cases*

These properties are presumably general, although they would certainly require adaptation in specific circumstances. When intricacy originates for instance from external collisions on a solid box, it must be carried by phonons and grows initially through phonon-phonon collisions (in which case, it moves again at the velocity of sound). In the case of intricacy for an electric signal in a conductor, the wave front will probably move at the Fermi velocity. When intricacy is carried by photons, the velocity of intricacy waves becomes $c$. Similar considerations would be true for long-range electromagnetic interactions (when the source is for instance an electric polarization in an ionized region). The front would move again at velocity $c$, but would have a large width, because there would be nonlinear term in the right-hand side of (5.3): The first term would become nonlocal because of retardation effects. The physics of intricacy could be therefore rich, although one will not try to push here its study farther.

Finally, one may notice that when the particle $M$ stops inside $S$., it continues to contribute as a source of intricacy but this source remains localized and its only relevant feature for the growth of intricacy is precisely this localization, so that nothing essential is changed in the discussion

## 6. Consequences of intricacy

The main conclusion of this work is the prediction of intricacy waves carrying entanglement, but its interpretation is rather unusual and raises some questions. The properties of locality, growth and transport of entanglement, as expressed in these waves, do not constitute phenomena in a proper sense, or at least a conventional one, since they do not reflect a behavior of some observables. They remain hidden in the past history of wave functions and are not expressed directly by the Schrödinger equation of evolution. One could wonder therefore whether they are not gratuitous constructions without real consequence.

A good reason for taking them seriously is the gain in understanding of entanglement, which results from its transport properties. When one looks on the other hand for real physical effects that would be sensitive to intricacy, only two fields seem open, eventually: They could be quantum computing and measurement theory, because both stand in an essential way on entanglement in the wave functions of a macroscopic system. To find whether intricacy could have consequences in these fields, one will consider first the relation of intricacy with decoherence.

*On decoherence and incoherence*

The discovery of decoherence came from measurement theory. One knew from Von Neumann and Schrödinger that this theory stands on the unitarity of a quantum evolution and on entanglement between a measured system $M$ and a measuring system $S$. As well known, this theory was also at least unsatisfactory [3]. Since then, the main substantial addition to this analysis came from decoherence theory [12, 13], which introduced the interaction between the measuring apparatus and its environment and showed a resulting break in unitarity in the $M + S$ system. There was however no drastic consequence for measurement theory, because decoherence did not affect the stability of the $M$-$S$ entanglement and it preserved particularly the values of channel probabilities. The main consequence of this conservation remained therefore, which is that never a unique datum would be expected to come out from an individual measurement. Decoherence is also the most preoccupying effect in the perspective of quantum computing and there is already a vast literature on this topic [14, 15].

Decoherence is a case of entanglement between two systems, except for two aspects. It does not deal with two systems $S$ and $S$' obeying together quantum laws, but one system $S$ is a measuring device whereas the second system is an environment $e$, which is not closed. One cannot therefore assert with certainty that the combined system $S+e$ obeys Schrödinger's equation. Moreover, the two systems $S$ and $e$ do not become separate after their interaction.

The standard theoretical description of decoherence brings it back however into the framework of entanglement, with a difference with Schrödinger's initial definition, namely that this entanglement holds between the measuring apparatus $S$ and elementary components in the environment $e$ (atoms or molecules or phonons...) rather than globally between the two systems $S$ and $e$ [16], just like was done for the $M$-$S$ entanglement in sections 2-5. Because of the short time of interaction between elementary components in $e$ with $S$, one can say that each interaction is of short duration and the separation between the system $S$ and an outside molecule, for instance, is rapid.

This prevalence of elementary interactions stressed another character of decoherence, which is its randomness. Most early studies of decoherence dealt with models of an incoherent environment in which the elementary components are described by random oscillators [17, 13, 18] or particles in the environment colliding randomly with a pointer in the apparatus [19]. The status of this randomness is not yet perfectly clear and received criticisms [20], but one will use it here now, before coming back to it in a moment.

The new point in which one is interested here, when decoherence is considered as an entanglement between a macroscopic system $S$ and the elements of its environment e, as in [16] for instance, is that it should also generate intricacy waves. To study this aspect, one will again consider $S$ as an atomic gas in a box and take $e$ as an atmosphere surrounding. $S$. Every collision of an external molecule on the box generates an entanglement between the inner state of $S$ and the outgoing state of the molecule. Inside $S$ this entanglement grows and moves as an intricacy wave, as was previously discussed. There is also an analogous effect in $e$ because the outgoing molecule collides with other molecules in the atmosphere and there is thus a contagion in $e$ of the corresponding intricacy with $S$. One will not deal however with this reciprocity of intricacy in the present paper and, to keep a clear simple view of the situation, one will consider mainly a model in which the atmospheric molecules do not interact together and only collide with the box (with transmission of the effect to the gas through phonons, as mentioned earlier).

A remarkable quantitative aspect of the situation is the very large number of intricacy waves in $S$. There are many collisions of molecules everywhere on the box, even during a small time interval, and one considers them usually as independent events. These independent collisions bring out as many distinct intricacies of $S$ with individual molecules in $e$. As shown previously, every individual collision of a molecule somewhere on the box generates an intricacy wave in the gas. This wave grows behind a spherical front, which moves with a velocity $v'$ and is centered on the point of impact of the molecule on the box. The number of the waves that are generated per unit time, is very large, of order $\tau_d^{-1} = n_e v_e L^2$ if $n_e$ denotes the number density of molecules in the environment, $v_e$ their average velocity and $L$ a typical length scale characterizing the box. ($\tau_d$ would be the rate of decoherence if the whole box acted as a pointer in a measuring device [19]). An intricacy wave, which moves at the sound velocity $v'$ spends a time of order $L/v'$ in the box before filling up this box completely and leaving after while a complete contagion of intricacy of the whole gas with the external molecule, which produced that wave. At any time, the number of the moving waves in the box is therefore of order $n_e (v_e/v') L^3$ and this is a tremendously large number, since it is of the same order of magnitude as the number of atoms in the box, when the gas in the box and the outside atmosphere are at comparable temperature and pressure.

This high disorder in the quantum state of $S$ does not affect however the probability of macroscopic observables in the gas and, in practice, intricacy waves contribute only to a general thermal disorder. It would seem therefore that they have no specific consequence, except when the quantity of interest is not an observable or an average value, but is concerned with a wave function itself. Said otherwise, a consequence of intricacy in the case of decoherence can only be expected in a quantum computation or a quantum measurement: Intricacy waves break down the unitary evolution of wave functions in $S$ and, since they are incoherent, these effects would be damaging in a quantum computer. Clever methods for correcting errors have been proposed to damp them [21], but one may wonder how an error corrections code in a large computer would work when incoherence is permanently brought in by many wave fronts and each front acts permanently as a source of incoherent perturbations within its width of order $\lambda$. .

Finally, one may come back to the assumption of incoherence among different external collisions. This is an old question and one will only call attention on a new feature, which directly linked with intricacy waves: Nothing macroscopic is isolated in the universe and an environment $e$ has always a super-environment $e$ ' with a boundary lying farther away. This distance is often used as an argument to say that a super-environment cannot influence an event occurring in $S$, because its effects take too long a time for $e'$ to be affected by an evolution in $S$ and react on it before its completion. The new element modifying this standard argument is now that the action of $e'$ on the state of $e$ and eventually on the state of $S$ proceeds through intricacy waves, which move permanently within $e$ and can bring their own incoherence into $S$ long after they left a distant part of the environment.

*Quantum measurements*

One will consider a simple case where a measurement consists in the interaction of a measuring system $S$ and a measured microscopic system $M$, where the state of $M$ is for instance initially a superposition $|s\rangle$ of two states $|1\rangle$ and $|2\rangle$, which are eigenvectors of an observable $Z$.:

$$|s\rangle = c_1|1\rangle + c_2|2\rangle. \tag{6.1}$$

The value of $Z$ controls the dynamics of $S$ and the final state of the system $S+M$ is accordingly strongly intricate [3].

One considers first the case when the system $S+M$ is isolated. In the simple case of a predictable measurement, when $c_1 = 1$ and $c_2 = 0$ for instance, the theory of the previous sections applies directly: There is a wave of intricacy of $S$ with the state $|1\rangle$ of $M$., which moves at a velocity $v'$ until complete contagion when the state of the apparatus is completely entangled with the state $|1\rangle$ of $M$. The same is true for state $|2\rangle$ when $c_1 = 0$ and $c_2 = 1$.

In the case of the superposition $|s\rangle$ in (6.1), the two intricacy waves are simultaneously present, with respective probabilities $p_j = |c_j|^2$ ($j = 1$ or 2), and they are completely independent if the system $S+M$ is isolated. A few points should be mentioned however for completeness: The theory in Section 3 must be enlarged to take account of two intricacies with the channels 1 and 2 and an intricacy index $q$ for the atoms in $S$ consists then in two parts $q_1$ and $q_2$. There is no interaction between an atomic state that is intricate with channel 1 and another state that is intricate with channel 2 and the operator $O_{nn'}$ in (3.7) becomes

$$P_{0n} \otimes P_{0n'} + P_{1n} \otimes P_{1n'} + P_{2n} \otimes P_{2n'}$$
$$+ S_{n1}P_{0n} \otimes P_{1n'} + P_{1n} \otimes S_{n'1}P_{0n'} + S_{n2}P_{0n} \otimes P_{2n'} + P_{2n} \otimes S_{n'2}P_{0n'}. \tag{6.2}$$

The field formulation in Section 4 involves then three fields $\varphi_0(x)$, $\varphi_1(x)$ and $\varphi_2(x)$, with no essential change otherwise in the discussion. The description of intricacy waves in Section 5 involves now there local probabilities of intricacy $f_0(x, t), f_1(x, t)$ and $f_2(x, t)$ such that $f_0 + f_1 + f_2 = 1$, whereas the evolution equation (5.3) becomes a set

$$\partial f_1 / \partial t = f_1 f_0 + (1/6)\nabla^2 f_1, \qquad \partial f_2 / \partial t = f_2 f_0 + (1/6)\nabla^2 f_2,$$
$$\partial f_0 / \partial t = -f_0(f_1 + f_2) + (1/6)\nabla^2 f_1, \qquad (6.3)$$

with a similar discussion of boundary conditions and wave fronts. Initial conditions and complete intricacy in a channel $j$ correspond then to $f_j(x,t) = p_j$ and there is no evolution in these channel probabilities $p_j$.

*Predecoherence and new possible effects in quantum measurements*

The most attractive consequence of entanglement locality is still conjectural, but especially worth a careful investigation. It deals with the possibility of fluctuations in the squared amplitudes $p_1$ and $p_2$ of the two measurement channels when incoherence, arising from intricacy with the environment, interferes with the growth of intricacy in the measurement channels. One will sketch only here the main idea in the case of a model where $M$ is a charged particle and $S$ a Geiger counter or a wire chamber. This detector contains an atomic gas, as in the previous examples, and there is an electric field in it. The basic detection steps, namely initial ionization by $M$ and secondary ionization cascades from the electric field can be worked out as in standard discussions by using the field methods of Section 4 and one takes this step as granted. The environment $e$ is again an outside atmosphere acting on a solid box, which encloses $S$.

One knows that decoherence, in the usual sense, cannot bring out fluctuations in $p_1$ and $p_2$, but this kind of decoherence acts only when the entanglement of $M$ and $S$ is complete or nearly complete, when ionization has become strong enough to yield a macroscopic polarization in $S$, which behaves like a pointer. One is interested on the contrary in the prior period when this intricacy is beginning, growing and moving. Intricacy with $e$, is permanently active and the associated intricacy waves are everywhere present and moving in $S$, bringing into it their own incoherence and breaking unitarity in the evolution of the system $M + S$. Because of this difference in timing, one may prefer to restrict the name "decoherence" to the damping of non-diagonal elements in the density matrix $\rho_{SM}$, after generation of a macroscopic signal, and speak of "predecoherence" for a prior effect, in spite of their common cause.

For definiteness, one will suppose that the charged particle $M$ follows two cleanly distinct trajectories in $S$, when its state is 1 or 2. As a matter of fact, the studies that have been made already work just as well when the time of entry of $M$ in $S$ differ or when channel 2 is mute, because $M$ does not hit $S$ in that channel. One will make no detailed study however in the present paper and only point out the main idea, which is simple.

Everything is a matter of intricacy, but of two different kinds. There is on one hand an entanglement of the measured system $M$ and the measuring apparatus $S$. The corresponding intricacies are initially produced through collisions of $M$ with atoms in $S$, producing excited atoms, ions and free electrons, which are intricate with the corresponding state of $M$. Later on, collisions of these intricate atoms with non-intricate ones increase intricacy, as discussed in previous sections.

There is also on the other hand an entanglement of the measuring apparatus $S$ and its environment $e$, as one described in the previous subsection. It proceeds through a large number of intricacy waves traveling in $S$ and one assumed that each one of these waves carries a random phase $\alpha$, because of randomness in external collisions. Ahead of the wave, no atom in $S$ is intricate with its source in $e$ and does not carry the phase $\alpha$. Behind the front of the wave, at some distance, every atom is on the contrary intricate with the source and carries the phase $\alpha$. As found in Section 5, the transition takes place near the wave front, in a region behind this front with a width of order $\lambda$, the mean free path of atoms.

The number of $\alpha$-phases traveling in $S$ is huge, of order $n_e(v_e/v') \, L^3$. The active part of a wave, where intricacy grows, has a width of order $\lambda$ however, and this means that at a point $x$ in the gas, there is always a number $N_w$ of active waves $w$ of order $n_e L^2 \lambda$ (the ratio of velocities being usually of the order of 1). For a gas of argon at standard conditions of temperature and pressure and a size $L$ of the detector of order 10 cm, there are about $10^{16}$ active waves at every place in $S$. This number looks tremendous, but its significance becomes more understandable when one remembers that these waves will be responsible later on for decoherence, which is a tremendously strong and rapid effect.

One arrives thus at the key point in the discussion, which deals with the coherent or incoherent character of atomic collisions. The two kinds of intricacy growth, between $S$ and $M$ on one hand and between $S$ and $e$ on the other hand, arise from the same atomic collisions. When a state of an atom $a$, which is for instance intricate with $|1\rangle$, collides with a state of an atom $a'$ the transition amplitude vanishes if the state of $a'$ is intricate with $|2\rangle$, because the Hamiltonian $H$ for the system $S+M$ as well as its extension $H'$ in Section 3 cannot induce a transition between the states $|1\rangle$ and $|2\rangle$. The interesting case occurs therefore when the state of the other $a'$ is neither intricate with $|1\rangle$ nor $|2\rangle$. The collision between $a$ and $a'$ carries a contagion of the intricacy of $a$ with $|1\rangle$ to the outgoing state of $a'$, but this transition is not always unitary. It is coherent and unitary if both states carry the same set of random phase $\alpha$ with $e$, otherwise it is incoherent. Although its probability of occurrence is still given by the square $|T|^2$ of a collision matrix element $T$, as in a coherent collision, unitarity (*i.e.,* phase continuity) is broken because of different random phases $\alpha$.

The previous counting of $N_w$ shows that most collisions entering in a contagion of intricacy with $|1\rangle$ are incoherent and this incoherence holds also for intricacy with $|2\rangle$ whereas the two processes are independent. Because of their incoherence, these collisions are not constrained by unitarity. Because of their independence, there is no obvious reason why they should conserve the channel probabilities $p_1$ and $p_2$, although the sum $p_1 + p_2$ remains equal to 1 since it is the trace of $\rho_{SM}$. Hence the question: are there fluctuations in $p_1$ and $p_2$, before they become frozen in macroscopic signals on which decoherence will act?
This question is obviously essential, since a positive answer could mean an interpretation of wave function collapse relying only on the quantum principles. This possibility was envisioned in a previous work [22], which would need deep revisions, corrections and adaptation in view of the present developments, but it showed however already that such a process, when made explicit, agrees with the non-separable character of quantum mechanics.

To conclude, one will first recall that unitarity and total entanglement are the essential impediments forbidding us to understand why there are unique data in quantum measurements, or why reality is unique. However, as one already knows that decoherence breaks down unitarity. What can be added now from the present work is that intricacy, *i.e.,* misunderstood or unnoticed local properties of entanglement, reveals where there could be a weak point in the long-received argument. This is still only a hint, of course, but it seems worth more study and is under more investigation, which will be published later.

**Figure caption**

Figure 1: An intricacy wave, The local probability of intricacy behind a wave front at x = 0, with a unit of distance taken as the atomic mean free path.

## References

[1] E. Schrödinger, '*Discussion of probability relations in separated systems*', Proc. Cambridge Phil. Soc. **31**, 555 (1935), **32**, 446 (1936)

[2] F. Laloë, '*Comprenons-nous vraiment la mécanique quantique?*' CNRS Editions, EDP Sciences, Paris 2011

[3] E. Schrödinger, '*Die gegenwärtige Situation in der Quantenmechanik*', Naturwissensschaften **23**, 807, 823, 844 (1935), reprinted in J.A Wheeler and W.H. Zurek, *Quantum mechanics and measurement*, Princeton University Press (1983)

[4] J. von Neumann, *Mathematische Grundlagen der Quantenmechanik*, Springer, Berlin (1932). English translation by R. T. Beyer, *Mathematical Foundations of Quantum Mechanics*, Princeton University Press (1955)

[5] M. B. Plenio, S. Virmani, *An introduction to entanglement measures*, Quant. Info. Comput. **7**, 1-51 (2005)

[6] N. Steenrod, *The topology of fiber bundles,* Princeton University Press (1951)

[7] M. L. Goldberger, K. M. Watson, *Collision Theory*, Wiley, New York (1964).

[8] L.S. Brown, *Quantum field theory*, Chapter 2, Cambridge University Press (1992)

[9] E. M Lifshitz, I. P Pitaevskii, *Physical kinetics*, Pergamon Press, oxford (1981)

[10] R. Dautray, J-L Lions, *Mathematical analysis and numerical methods for science and technology*. *Evolution problems, I, II*, Springer, Berlin, 2000

[11] L.D. Landau, E.M. Lifschitz, *Fluid mechanics*, Pergamon, London (1959)

[12] H. D. Zeh, *On the interpretation of measurement in quantum theory*, Found. Phys. **1**, 69-76 (1970)

[13] W. H. Zurek, *Environment-induced superselection rules*, Phys. Rev. **D 26**, 1862-1880 (1982)

[14] D. Mermin, *Quantum computer science, an introduction*, Cambridge University Press (2007)

[15] V. M. Akulin, A. Sarfati, G. Kurizki, S. Pellegrin, *Decoherence, entanglement and information protection in complex quantum systems*. Nato series II, Springer, Berlin (2005)

[16] E. Joos, H.D. Zeh, C. Kiefer, D. Giulini, K. Kupsch, I.O. Stamatescu, *Decoherence and the Appearance of a Classical World in Quantum Theory*, Springer, Berlin (2003)

[17] K. Hepp, E. H. Lieb, Helv. Phys. Acta, **46**, 573 (1973)

[18] A. O. Caldeira, A. J. Leggett, Physica A **121**, 587 (1983), Ann. Phys. (N. Y.) **149**, 374 (1983)

[19] E. Joos, H.D. Zeh, *The emergence of classical properties through interactions with the environment*, Z. Phys, **B 59**, 223-243 (1985)

[20] J.S. Bell, Helv. Phys. Acta, **48**, 93 (1975)

[21] P. W. Shor, *Scheme for reducing decoherence in quantum computer memory*, Phys; rev. **A 52**, 2493-96 (1995)

[22] R. Omnès, *Decoherence and wave function collapse*, Found. Phys. **41**, 1857-1880 (2011)